\documentclass[]{tMPH2e}
\usepackage{threeparttable}

\newcommand{\wn}{~\ensuremath{\mathrm{cm^{-1}}}}
\newcommand{\etal}{\emph{et al.}}

\newcommand{\bdpstate}{$B''\bar{B}^{1}\Sigma^{+}_{u}$}
\newcommand{\bdp}{$B''\bar{B}$}
\newcommand{\bpstate}{$B'^{1}\Sigma^{+}_{u}$}

\def\prl{Phys.\  Rev.\ Lett.\ }
\def\pra{Phys.\  Rev.\ A }

\def\jms{J. Mol.\ Spectrosc.\ }

\def\jcp{J.\ Chem.\ Phys. }
\def\molphys{Mol. Phys. }

\begin{document}
\title{VUV Spectroscopic Study of the \bdpstate\ State of H$_2$}

\author{G.D. Dickenson, W. Ubachs$^{\ast}$\thanks{$^\ast$Corresponding author. Email: w.m.g.ubachs@vu.nl}}
\maketitle
\begin{center}
\begin{scriptsize}
Department of Physics and Astronomy, LaserLaB, VU University, De Boelelaan 1081, 1081 HV Amsterdam, The Netherlands\\
\end{scriptsize}
\end{center}
\vspace{15pt}

\begin{abstract}
Spectral lines, probing rotational quantum states $J'=0,1,2$ of the inner well vibrations ($v' \leq 8$) in the \bdpstate\ state of molecular hydrogen, were recorded in high resolution using a vacuum ultraviolet (VUV) Fourier transform absorption spectrometer in the wavelength range 73-86 nm. Accurate line positions and predissociation widths are determined from a fit to the absorption spectra. Improved values for the line positions are obtained, while the predissociation widths agree well with previous investigations.
\end{abstract}

\section{Introduction}

The ($1s\sigma_s$, $4p\sigma_u$) \bdpstate\ state of H$_2$ is the third in a sequence of $np\sigma_u$ electronic excitations building states of $^1\Sigma_u^+$ symmetry. Due to strong interaction with the ($2s\sigma_s$, $2p\sigma_u$) doubly-excited state the \bdpstate\ state exhibits a double well structure in the adiabatic picture~\cite{Staszewska2002}.
Rovibrational levels in the inner well, referred to as $B''$, have been investigated by methods of classical spectroscopy several decades ago~\cite{Namioka1964,Monfils1965,Takezawa1970,Herzberg1972}. There it had been established that strong predissociation occurs, due to interaction with the continuum of the ($3p\sigma_u$) \bpstate\ state above the $n=2$ dissociation threshold. Only the $B'', v=0$ level, lying below the threshold, is unpredissociated. The $B''-X$ (2,0) band was investigated by direct one-photon laser excitation in the extreme ultraviolet~\cite{Rothschild1981}. Rovibrational levels in the outer well, referred to as $\bar{B}$, have been probed by multi-step laser excitation~\cite{deLange2001,Ekey2006} and are much longer lived.

The availability of vacuum ultraviolet (VUV) radiation from synchrotron sources has brought a revival of investigations of predissociative states in the benchmark H$_2$ molecule. In the studies at the BESSY II synchrotron the VUV is monochromatized before excitation, employing a bandwidth of 0.0012 nm (or 1.8 \wn) in the spectroscopic studies. This setup is particularly suitable to investigate the dynamics and quantitative decay rates of predissociative states in H$_2$ \cite{Glass-Maujean2007b,Glass-Maujean2010a,Glass-Maujean2010b}, including the $B''^1\Sigma_u^+$ state~\cite{Glass-Maujean2007a}, and a series of P(1) lines probing high-lying vibronic states of $^1\Sigma_u^+$ symmetry~\cite{Glass-Maujean2011}. Alternatively, the broadband VUV output of the SOLEIL synchrotron is fed into a Fourier-transform spectrometer (FTS) to obtain high resolution absorption spectra. This setup was previously applied in the investigation of bound \cite{Ivanov2010} and predissociative states in molecular hydrogen \cite{Dickenson2010a,Dickenson2011}. Results from the latter setup on the \bdpstate\ state for vibrational levels $v' \leq 8$ are reported in the present study. They are compared with results from previous experiments, as well as with theoretical studies.

\section{Experimental}

Absorption spectra were recorded with the VUV-FTS at the DESIRS beamline of the synchrotron SOLEIL. The operation of the instrument has been described previously~\cite{deOliveira2009,deOliveira2011}, as well as its application to the high resolution spectroscopy of the hydrogen molecule~\cite{Ivanov2010,Dickenson2010a,Dickenson2011}.
In short, broadband synchrotron radiation is passed through a t-shaped absorption cell of length 100 mm and inside diameter of 12 mm. The cell contains H$_{2}$ gas flowing under quasi-static conditions. The t-shaped cell is covered by a second cell which is used as a liquid nitrogen reservoir, thereby cooling the H$_{2}$ sample to temperatures as low as 100 K.

The undulator-based synchrotron radiation spans $\sim$5000 \wn\ and measurements are split into three overlapping windows, covering the range 73 - 86 nm. A variety of pressures are used to record the \bdp-$X$ transitions under unsaturated conditions. The wavelength accuracy, with the intrinsic calibration of the FTS instrument, is as good as 0.1 \wn\ for the narrow lines, increasing to 0.3 \wn\ for the broadest predissociated lines. The resolution is determined by a number of settings on the FTS instrument~\cite{deOliveira2011}. For the measurements presented here the settings were adjusted to produce an instrument profile (sinc function) with a FWHM of 0.35 \wn. The fitting procedures to deduce peak positions of the resonances as well as predissociation widths are documented in Ref.~\cite{Dickenson2011}. An essential ingredient in the convolution is the Gaussian width  of 0.6 \wn, representing the Doppler width at the ambient temperatures. The non-linear absorption depth is accounted for using the Beer-Lambert law and the resultant profiles are deconvolved from the instrument function.

\section{Results}

\begin{figure}
\centering
\includegraphics[width=0.75\linewidth]{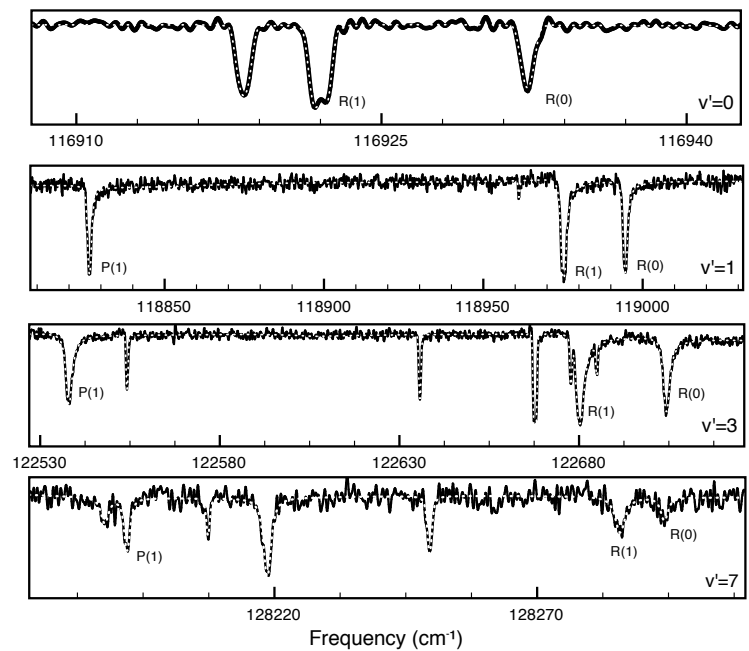}
\caption{Sections of the FTS-VUV absorption spectra depicting from top to bottom lines in the (0,0), (1,0), (3,0) and (7,0) bands of the $B''-X$ system. The dotted white lines represent a fit of the data with the convolved functions. Note the different frequency scale for the (0,0) band in the upper panel showing the unpredissociated lines.}
\label{fig:Profiles}
\end{figure}

\begin{table}
\caption{Transition frequencies for the R(0), R(1) and P(1) lines exciting the \bdpstate\ state for vibrations up to $v'=8$. $\Delta$ represents the present measurements minus the previous results as determined by Monfils~\cite{Monfils1965} for vibrations $v'=0-5$ (except for the P(1) line for $v'=5$), by Glass-Maujean \etal\ \cite{Glass-Maujean2007a} for the R(0) and R(1) lines for vibrations $v'=6-8$, and by Glass-Maujean \etal\ \cite{Glass-Maujean2011} for the P(1) lines for $v'=5-8$. All values in \wn.}
\label{Tab:Uncertainty}
\begin{threeparttable}
\begin{tabular}{c r@{.}l r@{.}l r@{.}l c r@{.}l r@{.}l r@{.}l}
\\
\hline
\hline
\multicolumn{1}{c}{Transition}&\multicolumn{2}{c}{Frequency}&\multicolumn{2}{c}{$\Delta$}&\multicolumn{2}{c}{$\Gamma$}&\multicolumn{1}{c}{Transition}&\multicolumn{2}{c}{Frequency}&\multicolumn{2}{c}{$\Delta$}&\multicolumn{2}{c}{$\Gamma$}\\
\hline
\hline\\
\multicolumn{7}{c}{\bdp (0,0)}&\multicolumn{7}{c}{\bdp (1,0)}\\[1mm]
R(0)&116932&13&0&58&\multicolumn{2}{c}{--}&R(0)&118994&41&0&55&0&43\\
R(1)&116921&93&1&87&\multicolumn{2}{c}{--}&R(1)&118975&11&0&27&0&10\\
P(1)&116767&57&0&72&\multicolumn{2}{c}{--}&P(1)&118826&29&0&75&0&54\\[1mm]
\multicolumn{7}{c}{\bdp (2,0)}&\multicolumn{7}{c}{\bdp (3,0)}\\[1mm]
R(0)\tnote{s}&120917&54&0&11&1&87&R(0)&122704&01&0&16&1&25\\
R(1)&120895&56&-0&39&0&77&R(1)&122680&11&0&30&1&12\\
P(1)&120750&95&1&05&0&95&P(1)&122537&80&0&39&1&53\\[1mm]
\multicolumn{7}{c}{\bdp (4,0)}&\multicolumn{7}{c}{\bdp (5,0)}\\[1mm]
R(0)&124354&46&0&81&1&16&R(0)&125856&77\tnote{t}&-1&30&0&77\\
R(1)&124332&91&-1&44&1&11&R(1)\tnote{b}&125851&72\tnote{t}&-0&79&3&49\\
P(1)&124189&19&0&10&1&17&P(1)&125730&01\tnote{t}&-0&09&1&11\\[1mm]
\multicolumn{7}{c}{\bdp (6,0)}&\multicolumn{7}{c}{\bdp (7,0)}\\[1mm]
R(0)\tnote{b}&127129&65&-1&57&0&97&R(0)&128338&72&0&31&1&48\\
R(1)&127082&90&-6&12&3&14&R(1)&128285&74&-3&27&2&09\\
P(1)&126976&05&0&35&2&29&P(1)&128191&91\tnote{t}&+4&11&1&02\\[1mm]
\multicolumn{7}{c}{\bdp (8,0)}\\[1mm]
R(0)&129436&64&-4&82&2&57\\
R(1)\tnote{b}&129393&60&-2&64&4&04\\
P(1)&129281&35&0&35&2&55\\[1mm]
\hline
\hline
\end{tabular}
\begin{tablenotes}
\footnotesize
\item[s] Saturated.
\item[b] Blended.
\item[t] Tentative
\end{tablenotes}
\end{threeparttable}
\end{table}

P(1), R(0) and R(1) spectral lines in the $B''-X$ ($v',0$) bands for $v' \leq 8$ were observed in the wavelength range 73-86 nm. Fig.~\ref{fig:Profiles} shows typical spectra of the unpredissociated lines of the (0,0) band and the predissociated lines in the (1,0) and (3,0) bands. Numerical analysis of the spectra resulted in the values for the line positions and the predissociation widths $\Gamma$ presented in Table~\ref{Tab:Uncertainty}.
The R(0) line for the (2,0) band was saturated in all measurements. Some of the lines analyzed were blended by other transitions in the H$_2$ absorption spectrum; values were obtained via deconvolution. The R(0) line in the (6,0) band is rather heavily blended with the Q(1) line of the $D-X$ (8,0) band, which was not noticed in previous studies of the $D^1\Pi_u$ state~\cite{Glass-Maujean2007a,Dickenson2010a}. Inspection of the 100 K static gas spectrum, as well as previously analyzed jet-cooled and liquid-He cooled static gas spectra~\cite{Dickenson2010a}, clearly show partially overlapping resonances consisting of a narrow component and a weaker asymmetric component at the blue side. The first is assigned as the $D-X$ (8,0) Q(1) line and the second to the $B"-X$ (6,0) R(0) line, with a transition frequency obtained from deconvolution.

In the Table the line positions are compared with experimental values from literature. The classical spectroscopic data of Monfils~\cite{Monfils1965}, available for vibrational states \bdpstate, $v'=0-5$, are in agreement within 1 \wn, except for the P(1) line of the (5,0) band. For the bands $B''-X$ (6,0) to (8,0) a comparison was made with the experimental data for the R(0) and R(1) lines of Glass-Meaujean \emph{et al.}~\cite{Glass-Maujean2007a}, and for the P(1) lines from a later study by Glass-Maujean \emph{et al.}~\cite{Glass-Maujean2011}. These data yield somewhat less agreement, reflecting the lower absolute accuracy of the wavelength calibration of the BESSY II spectrometer.

The assignment of the resonances pertaining to the $B''-X$ system was straightforward, searching for broadened lines and following previous studies. In the assignment procedure a distinction was made between strongly broadened lines of the $B''-X$ and $D-X$ systems, where the lines in the $D-X$ system exhibit decisively asymmetric Fano-lineshapes, where lines of the $B''-X$ system are nearly symmetric; therefore in the present study no Fano $q$-parameters were derived.

The assignment of the lines in the $B''-X$ (5,0) band had caused difficulties in the past. Where Monfils~\cite{Monfils1965} had identified the P(1) line at $125709.61$ \wn, Takezawa~\cite{Takezawa1970} identified it at $125683.4$ \wn. At the latter position the $B''-X$ (5,0) R(1) line was identified by Herzberg and Jungen~\cite{Herzberg1972}. A firm reassignment was based on the early MQDT calculations by Jungen and Atabek~\cite{Jungen1977}, putting the P(1) line at $125737.2$ \wn. These MQDT calculations were later updated~\cite{Glass-Maujean2011}, putting the P(1) line just 0.7 \wn\ higher than a line identified at $125730.1$  \wn. For the assignment of the R(1) line in the (5,0) band we follow Ref.~\cite{Glass-Maujean2011}, although this line is 45 \wn\ separated from a simplified calculation based on a direct solution of the Schr\"{o}dinger equation. For the R(0) line we follow the assignment by Monfils~\cite{Monfils1965}, since no updated values are reported. For these reasons the assignment of R(0) and R(1) lines are indicated as tentative in Table~\ref{Tab:Uncertainty}.

The assignment of the P(1) line in the (7,0) band is tentative. As shown in the the lower panel of Fig.~\ref{fig:Profiles} a strong line
at 128\,219 \wn\ might also qualify for this P(1) line, but the present assignment agrees with previous assignments.
Moreover it agrees reasonably with the MQDT calculations of Jungen and Atabek, which is considered the most reliable basis for assigning lines.
Note the additional line at 128\,188 \wn, which in a lower resolution study may be overlapped with the presently assigned P(1) line.

\vspace{0.5cm}

\begin{figure}
\centering
\includegraphics[width=0.75\linewidth]{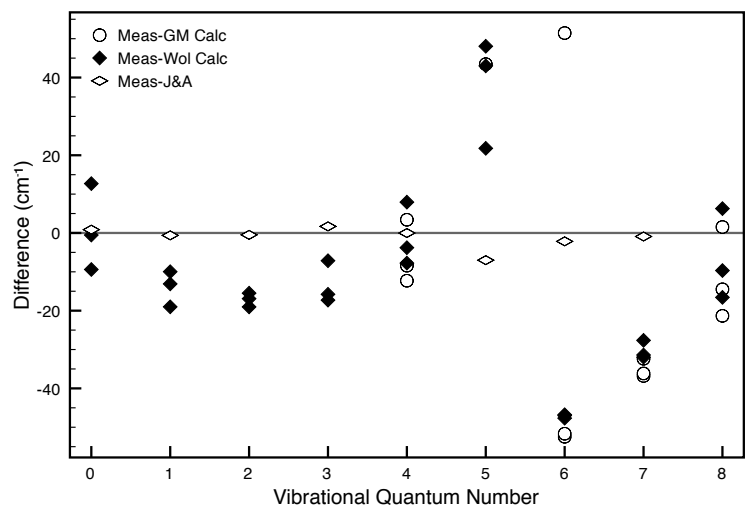}
\caption{Comparison between the present experimental values for the level energies of \bdpstate, $v', J'$ states and existing calculations. "J\&A" refers to calculations based upon the MQDT framework of Jungen and Atabek~\cite{Jungen1977} (for $J'=0$ levels only); "Wol" refers to \emph{ab initio} calculations including non-adiabatic effects by Wolniewicz~\cite{WolniewiczPrivComm}; "GM" refers to direct solutions of the Schr\"{o}dinger equation~\cite{Glass-Maujean2007a}. Various data points for the same $v'$ level pertain to $J'=0$, 1 and 2.
}
\label{fig:Theory}
\end{figure}

The present accurate determinations of transition frequencies in the $B''-X$ system are compared to calculations of the level energies in the \bdpstate\ state in Fig.~\ref{fig:Theory}. The MQDT calculations from Jungen and Atabek performed some three decades ago~\cite{Jungen1977}, still stand as an accurate basis for assignment. These MQDT calculations were recently updated later~\cite{Glass-Maujean2011} providing even better agreement with observation. Unfortunately the MQDT calculations were only performed for $J=0$ levels, where rotational-electronic $^1\Pi_u$ $\sim$ $^1\Sigma_u^+$ interactions have no influence. Alternatively Wolniewicz performed \emph{ab initio} calculations including non-adiabatic effects. At the time of that calculation only level energies for the HD isotopomer were explicitly reported~\cite{Reinhold1999}, as well as level energies for the $\bar{B}$ state of H$_2$~\cite{deLange2001}, but the $B''$ inner well levels for H$_2$ were calculated as well~\cite{WolniewiczPrivComm} and are now included in Fig.~\ref{fig:Theory}. The calculations by Glass-Meaujean \emph{et al.}~\cite{Glass-Maujean2007a} represent direct solutions of the Schr\"{o}dinger equation based on the adiabatic potential of Ref.~\cite{Reinhold1999}, ignoring non-adiabatic effects.

\begin{figure}
\centering
\includegraphics[width=0.75\linewidth]{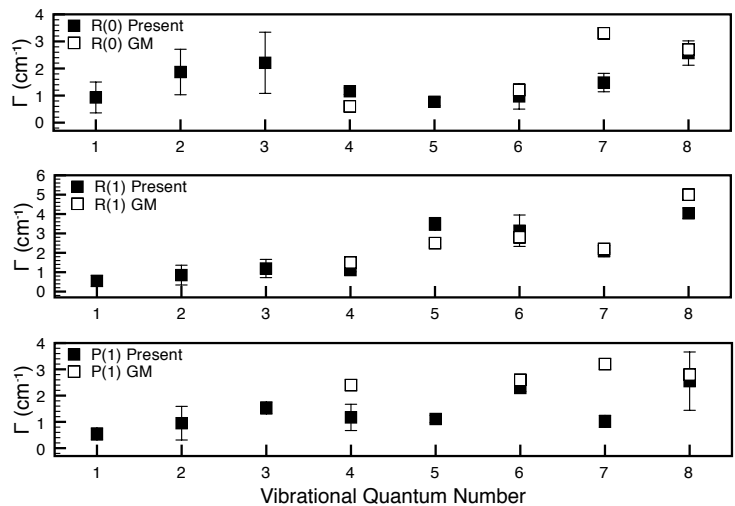}
\caption{Predissociation widths for \bdpstate\ levels for $J=0-2$ levels for $v'=1$ to 8 in comparison with the measurements by Glass-Meaujean \emph{et al.}~\cite{Glass-Maujean2007a} (only for $v' \geq 4$).}
\label{fig:Widths}
\end{figure}

\vspace{0.5cm}
The predissociation widths $\Gamma$, associated with the natural lifetimes of the excited states were deconvoluted from the observed line profiles by procedures described previously~\cite{Dickenson2010a,Dickenson2011}. The data are plotted in Fig.~\ref{fig:Widths} for comparison with the previous data from Ref.~\cite{Glass-Maujean2007a}, where values were obtained for $v' \geq 4$. Good agreement is found throughout, except for the R(0) and P(1) lines in the (7,0) band, where a large disagreement is found. For the P(1) line this may be caused by the nearby lying line at 128\,188 \wn, which may not be resolved in the lower resolution study at BESSY II~\cite{Glass-Maujean2007a}.

\section{Conclusion}

The present study reports the most accurate assignment of lines in the $B''-X$ system. Line positions are measured to an accuracy of 0.1 \wn.
The experimental data compare favorably with theoretical data from MQDT calculations, first the early calculations by Jungen and Atabek~\cite{Jungen1977} and later refined calculations by Glass-Maujean \emph{et al.}~\cite{Glass-Maujean2011}.  The MQDT calculations provide a better representation of the level energies of H$_2$ in this energy range than the direct treatment solving Schr\"{o}dinger equation~\cite{Glass-Maujean2007a}, even if non-adiabatic effects are explicitly included~\cite{Reinhold1999,WolniewiczPrivComm}. Unfortunately the MQDT calculations for the \bdpstate\ have not been extended to $J>0$ levels. Hence, the definite assignment of the lines in the (5,0) band await $J$-dependent MQDT calculations.

\section*{Acknowledgment}

The authors are grateful to the SOLEIL staff scientists L.
Nahon, N. de Oliveira and D. Joyeux for the hospitality and
for the collaboration. This work
was supported by the Netherlands Foundation for Fundamental
Research of Matter (FOM).


\begin{thebibliography}{}

\bibitem{Staszewska2002} G. Staszewska and L. Wolniewicz,
\jms \textbf{212}, 208 (2002).

\bibitem{Namioka1964}
T. Namioka,
\jcp \textbf{41}, 2141 (1964).

\bibitem{Monfils1965}
A. Monfils,
\jms \textbf{15}, 265 (1965).

\bibitem{Takezawa1970}
S. Takezawa,
\jcp \textbf{52}, 2575 (1970).

\bibitem{Herzberg1972}
G. Herzberg and Ch. Jungen,
\jms \textbf{41}, 425 (1972).

\bibitem{Rothschild1981}
M. Rotschild, H. Egger, R. T. Hawkins, J. Bokor, H. Pummer, and C. K. Rhodes,
\pra \textbf{23}, 206 (1981).

\bibitem{deLange2001}
A. de Lange, W. Hogervorst, W. Ubachs, and L. Wolniewicz,
\prl \textbf{86}, 2988 (2001).

\bibitem{Ekey2006}
R. C. Ekey Jr., A. Marks, and E. F. McCormack,
\pra \textbf{73}, 023412 (2006).

\bibitem{Glass-Maujean2007b}
M. Glass-Maujean, S. Klumpp, L. Werner, A. Ehresmann and H. Schmoranzer,
\molphys \textbf{105}, 1535 (2007).

\bibitem{Glass-Maujean2010a}
M. Glass-Maujean, Ch. Jungen, H. Schmoranzer, A. Knie, I. Haar, R. Hentges, W. Kielich, K. J\"{a}nk\"{a}l\"{a}, and A. Ehresmann,
\prl \textbf{104}, 183002 (2010).

\bibitem{Glass-Maujean2010b}
M. Glass-Maujean, Ch. Jungen, G. Reichardt, A. Balzer, H. Schmoranzer, A. Ehresmann, I. Haar, and P. Reiss,
\pra \textbf{82}, 062511 (2010).

\bibitem{Glass-Maujean2007a}
M. Glass-Maujean, S. Klumpp, L. Werner, A. Ehresmann and H. Schmoranzer,
\jcp \textbf{126}, 144303 (2007).

\bibitem{Glass-Maujean2011}
M. Glass-Maujean, Ch. Jungen, H. Schmoranzer, I. Haar, A. Knie, P. Reiss, and A. Ehresmann,
\jcp \textbf{135}, 144302 (2011).

\bibitem{Ivanov2010}
T.I. Ivanov, G.D. Dickenson, M. Roudjane, N. de Oliveira, D. Joyeux, L. Nahon, W-\"{U}. L. Tchang-Brillet, and W. Ubachs,
\molphys \textbf{108}, 771 (2010).

\bibitem{Dickenson2010a}
G. D. Dickenson, T. I. Ivanov, M. Roudjane, N. de Oliveira, D. Joyeux, L. Nahon, W-\"{U} L. Tchang-Brillet, M. Glass-Maujean, I. Haar, A. Ehresmann,  and W. Ubachs,
\jcp \textbf{133}, 144317 (2010).

\bibitem{Dickenson2011}
G.D. Dickenson, T. I. Ivanov, W. Ubachs, M. Roudjane, N. de Oliveira, D. Joyeux, L. Nahon,  W -\"{U}. L. Tchang-Brillet, M. Glass-Maujean, H. Schmoranzer, A. Knie, S. Kuebler, and A. Ehresmann,
\molphys \textbf{109}, 2693 (2011).

\bibitem{Jungen1977}
Ch. Jungen and O. Atabek,
\jcp \textbf{66}, 5584 (1977).

\bibitem{Reinhold1999}
E. Reinhold, W. Hogervorst, W. Ubachs, and L. Wolniewicz,
\pra \textbf{60}, 1258 (1999).

\bibitem{WolniewiczPrivComm}
L. Wolniewicz, Private communication (1999). In Ref.~\cite{Reinhold1999} the methods for the calculation
were presented but results only explicitly given for the HD molecule.


\bibitem{deOliveira2009}
N. de Oliveira, D. Joyeux, D. Phalippou,  J. C. Rodier, F. Polack, M. Vervloet, and L. Nahon,
Rev. Sci. Inst. \textbf{80}, 043101 (2009).

\bibitem{deOliveira2011}
N. de Oliveira, M. Roudjane, D. Joyeux, D. Phalippou,  J. C. Rodier, and L. Nahon,
Nat. Photon \textbf{5}, 149 (2011).





\end{thebibliography}
\end{document}